\documentstyle[epsf,12pt]{article}

\newcommand{\z}{&\hspace*{-8pt}}

\newcommand{\ep}{\varepsilon}
\newcommand{\Li}[2]{{\mbox{Li}}_{#1\!}\left(#2\right)}
\newcommand{\Ls}[2]{{\mbox{Ls}}_{#1}\!\left(#2\right)}
\newcommand{\LS}[3]{{\mbox{Ls}}_{#1}^{(#2)}\!\left(#3\right)}
\newcommand{\s}[1]{\sum_{n=1}^\infty
  \frac{1}{\left( 2n \atop n\right)} \frac{1}{n^#1}}

\newcommand{\Fh}[2]{{}_{#1}F_{#2}}
\newcommand{\Fs}[3]{\left(\begin{array}{c}#1\,; \\#2\,;\end{array}#3\right)}

\begin{document}
\begin{center}

\vspace{15mm}

{\large \bf  Single-scale diagrams and multiple binomial sums.}

\vspace{15mm}

{\large
M.~Yu.~Kalmykov%
\footnote{~E-mail: kalmykov@ifh.de\\
On leave of absence from BLTP, JINR, 141980, Dubna (Moscow Region), Russia},~
O.~Veretin%
\footnote{~E-mail: veretin@ifh.de}
}

\vspace{5mm}

{\it ~DESY Zeuthen, Theory Group, Platanenallee 6, D-15738, Zeuthen, Germany}

\vspace{15mm}

\end{center}

\begin{abstract}
The $\ep$-expansion of several two-loop self-energy diagrams with
different thresholds and one mass 
are calculated. On-shell results are reduced to multiple binomial 
sums which values are presented in analytical form. 
\end{abstract}


\vspace*{10pt}

{\it PACS number(s)}: 12.38.Bx

\thispagestyle{empty}
\setcounter{page}0
\newpage

\section{Introduction}

Substantial progress in multiloop Feynman diagram calculations
in recent years requires computation of scalar master integrals. 
Often the problem involving different mass scales can
be reduced (e.g. by expanding) to integrals depending only 
on a single scale.
Thus single-scale diagrams 
(e.g. bubbles with one non-zero mass, massless self-energy,
massive on-shell self-energy integrals, etc.) form an important class of Feynman diagrams.
Such integrals arise for example in renormalization group calculations.
The structure of massless integrals is well understood now 
\cite{massless1,massless2}. 
In particular recently a correspondence
between knot theory and massless diagrams \cite{knot} has been found which
can serve as a very useful guide to find the transcendental numbers
which occur with rational coefficient in the counterterms.
This relationship is known only for diagrams that are free of subdivergences.

  The transcendental structure of massive single-scale diagrams 
is less investigated \cite{cl2}-\cite{master3} (see also \cite{cln}, \cite{kl}).
In particular, we do not know whether there exist a theory to predict the transcendental 
numbers for these diagrams.
Recently it was observed \cite{master2} that all two-loop 
massive on-shell diagrams of propagator type without subdivergences can be
written in following way
\begin{equation}
m^2 {\bf I}_0 \left.\right|_{p^2=m^2} = 
   r_1 \zeta_3 + r_2 \pi \Ls{2}{\frac{\pi}{3}} + r_3 i \pi \zeta_2 
     + {\cal O}(\varepsilon),
\label{first}
\end{equation}
where $\zeta_a=\zeta(a)$ is the Riemann $\zeta$-function, 
$r_j$ are rational coefficients and definition of $\Ls{n}{z}$ is 
given by (\ref{ls}). 
This observation suggests a conjecture that irrationalities occurring
in these diagrams are defined by the topology of a diagram
but not e.g. by the distribution of the masses on lines. 
In this paper we test this conjecture in the next order of
the $\varepsilon$-expansion.  

Another problem under consideration is the test of the hypothesis about
the connection between transcendental numbers occurring in the $\ep$-expansion
of diagrams and the presence of certain massive-particles-cuts. This conjecture 
reads as follows: zero-, one- and three massive particle cuts 
give rise to appearance of structures $\pi^j \zeta_m (\ln 2)^n \Li{p}{1/2}$,
where ${\rm Li}_p(x)$ is polylogarithm, 
or more complicated structures associated with Euler--Zagier sums 
(or multidimensional zeta/harmonic sums) 
\cite{massless2, knot, Broadhurst3, kk, mathematics}
\begin{eqnarray}
\zeta(a_1,\ldots,a_k)=\sum_{n_i>n_{i+1}}\prod_{j=1}^{k}
\frac{({\rm sign}\,a_j)^{n_i}}{n_i^{|\,a_j|}},
\label{euler}
\end{eqnarray}
whereas the two massive particle particle cuts bring other 
transcendental 
numbers connected with ``sixth root of unity'' \cite{Broadhurst3,master3}:
$\left( \frac{\pi}{\sqrt{3}} \right)^k \zeta_m (\ln 3)^n \Ls{p}{z_i} \LS{q}{r}{z_j}$,
where $z_k= \left\{ \frac{\pi}{3}, \frac{2\pi}{3} \right\}$ and 
$\Ls{n}{z}$ and $\LS{n}{m}{z}$ are so-called log-sine integrals
\cite{Lewin} defined by
\begin{equation}
\label{ls}
\Ls{n}{\theta}  =  - \int\limits_0^\theta
      \ln^{n-1} \left| 2\sin\frac{\phi}{2}\right| \, d\phi, 
~~~~~~
\LS{n}{m}{\theta} =  - \int\limits_0^\theta
   \phi^m \ln^{n-m-1} \left| 2\sin\frac{\phi}{2}\right| \, d\phi.
\end{equation}
We will show here that some of these irrationalities are related to
sums \cite{sum0}-\cite{sum2}
\begin{equation}
\sum_{n=1}^\infty \frac{1}{\left( 2n  \atop n \right)} \frac{1}{n^c}
\prod_{a,b,i,j}
\left[ \sum_{m=1}^{n-1} \frac{1}{m^a} \right]^i
\left[ \sum_{k=1}^{2n-1} \frac{1}{k^b} \right]^j
\label{binsum}
\end{equation}
which we call {\it multiple binomial sums}. 

  The question about transcendental structures connected 
with four- and more massive particle cuts remains open.

\section{Results}

\begin{figure}[bth]
\centerline{\vbox{\epsfysize=30mm \epsfbox{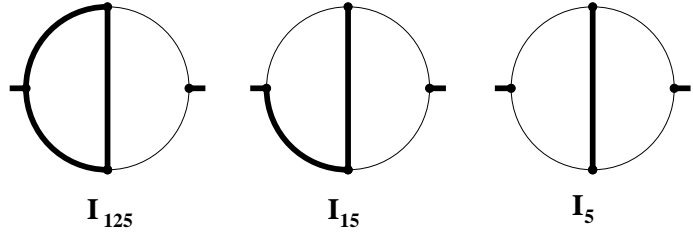}}}
\caption{Bold and thin lines correspond to massive and
massless propagators, respectively.}
\label{mixing}
\end{figure}

  As examples we consider the diagrams shown in Fig.\ref{mixing}. 
All diagrams posses different cuts in the external variable $p^2$
with following values of thresholds 
$I_{125} = \{ 0,4m^2 \}$,  
$I_{15}  = \{0, 1m^2, 4m^2\}$,  
and 
$I_{5}   = \{0, 1m^2\}$.  

  To evaluate these diagrams we use the semianalytic method developed 
in Ref. \cite{semianalytic}. This approach is based on a possibility to
restore analytical results in terms of harmonic sums
from several first coefficients of the small momentum 
expansion \cite{asymptotic}. In Ref. \cite{semianalytic} the ${\cal O}(1)$
parts of the diagrams shown in Fig.\ref{mixing} were found 
\footnote{The finite part of ${\bf I_5}$ is given in \cite{st}, 
${\bf I_{125}}$ in \cite{broadhurst1} 
and exact result for ${\bf I_{125} }$
in terms of hypergeometric function presented in \cite{hyper3}.}.
Here we extend these results calculating their $\ep$-parts. Omitting
all technical details that can be found in the above paper we present 
the results of our calculation%
\footnote{
We are working in Minkowski
space-time with dimension $N=4-2\ep$. 
For each loop we assume a common normalization factor 
$(m^2 e^\gamma)^\ep/\pi^\frac{N}{2}$, where $\gamma$ is Euler constant. 
}.
We find ($z=p^2/m^2$)
%
\begin{eqnarray}
&& {\bf I_{125}}  = 
\frac{1}{p^2} \sum_{n=1}^\infty \frac{1}{\left( 2n \atop n\right)} z^n
\Biggl[  
\frac{-\ln(-z) }{n^2} + \frac{3}{n^3}
+ \varepsilon \Biggl(
\frac{\ln^2 (-z)}{2 n^2}
- \ln (-z) \left\{\frac{2}{n^2} + \frac{S_1(n-1)}{n^2}
\right\}
\nonumber \\ && 
- \frac{1}{n^4} + \frac{6}{n^3} - \frac{\zeta_2}{n^2}
+ \frac{11}{n^3} S_1(n-1)  - \frac{4}{n^3}  S_1(2n-1)  
\Biggr)
+ {\cal O}(\varepsilon^2)
\Biggr],
\label{eq1}
\end{eqnarray}
%
%
\begin{eqnarray}
&& {\bf I_{15}}  = 
\frac{1}{p^2} \sum_{n=1}^\infty z^n
\Biggl[- \frac{\ln(-z) }{n^2} - \frac{\zeta_2}{n}
+ \frac{2}{n^3}  + \frac{3}{n} V_2(n-1)  
 + \frac{1}{\left( 2n\atop n\right)} \frac{4}{n^3} 
\nonumber \\ && 
+ \varepsilon \Biggl(
\frac{\ln^2 (-z)}{2 n^2}
-\ln (-z) \left\{\frac{1}{n^3} + \frac{2}{n^2} + \frac{2}{n^2} S_1(n-1)
\right\}
+ \frac{\zeta_3}{n} 
- \frac{2}{n} \zeta_2 
+ \frac{6}{n^3} S_1(n-1)  
\nonumber \\ && 
- \frac{3}{n^2}  V_2(n-1)  
+ \frac{3}{n} V_3(n-1)
+ \frac{15}{n} V_{2,1}(n-1)
- \frac{6}{n} \tilde{V}_{2,1}(n-1)
+ \frac{4}{n^3}
+ \frac{6}{n} V_2 (n-1)
\nonumber \\&& 
+ \frac{1}{\left( 2n \atop n\right)}
\left\{ 
 \frac{8}{n^3}
+ \frac{20}{n^3} S_1(n-1)
- \frac{8}{n^3} S_1(2n-1)
\right\} 
\Biggr)
+ {\cal O}(\varepsilon^2)
\Biggr],
\label{eq2}
\end{eqnarray}
%
%
\begin{eqnarray}
&& {\bf I_5}  = 
\frac{1}{p^2} \sum_{n=1}^\infty (-z)^n
\Biggl[
-\frac{\ln^2(-z)}{n}  + \frac{2}{n^2} \ln (-z)
- \frac{2}{n} \zeta_2 + \frac{4}{n} K_2(n-1)
- \frac{2}{n^3} - 2 \frac{(-)^n}{n^3}
\nonumber \\ && 
+ \varepsilon \Biggl(
\frac{\ln^3(-z)}{n}
- \frac{\ln^2(-z)}{n} \left\{ \frac{2}{n} + \frac{3}{n^2} + \frac{S_1(n-1)}{n} 
\right \}
+ \ln (-z) \left\{ \frac{4}{n^2} + \frac{6}{n^3} 
+ \frac{2}{n^2} S_1(n-1) \right. 
\nonumber \\ && \left.
+ \frac{2}{n} S_2(n-1) - \frac{2}{n} \zeta_2 \right \}
+ \frac{2}{n} \zeta_3 
+ \zeta_2 \left\{ \frac{2}{n^2} - \frac{4}{n} - \frac{2}{n} S_1(n-1)
\right \}
- \frac{4}{n^3} - \frac{6}{n^4} - \frac{2}{n^3} S_1(n-1)
\nonumber \\ && 
- \frac{2}{n^2} S_2(n-1) - \frac{2}{n} S_3(n-1)
+ \frac{8}{n} K_2(n-1)
+ \frac{12}{n} K_{2,1}(n-1)
+ \frac{4}{n} K_2(n-1) S_1(n-1)
\nonumber \\ && 
+ \frac{2}{n} K_3(n-1)
- (-)^n \left\{
\frac{4 }{n^3}
+ \frac{2 }{n^4}
+ \frac{8 }{n^3} S_1(n-1)
\right\} \Biggr)
+ {\cal O}(\varepsilon^2)
\Biggr], 
\label{eq3}
\end{eqnarray}
where we use the following notations for the finite sums elaborated 
in \cite{semianalytic}
$$
S_a(n) =  \sum_{j=1}^{n} \frac{1}{j^a},
~~~~K_a(n) =  -\sum_{j=1}^{n} \frac{(-)^j}{j^a},
~~~~ K_{a,b}(n) =  -\sum_{j=1}^{n} \frac{(-1)^j}{j^a} S_b(j-1),
$$
$$
V_a(n)  =  \sum_{j=1}^{n} {2j\choose j}^{\!\!\!-1} \frac{1}{j^a},
~~~~ V_{a,b}(n)  =  \sum_{j=1}^{n}
           {2j\choose j}^{\!\!\!-1} \frac{1}{j^a} S_b(j-1),
~~~~\tilde{V}_{a,b}(n) = \sum_{j=1}^{n}
      {2j\choose j}^{\!\!\!-1} \frac{1}{j^a} S_b(2j-1).
$$
Sums $K_a$, $K_{a,b}$ were also used in \cite{kk}.
Sums $V_a$ and $\tilde{V}_a$ were predicted by differential equation method
\cite{kotikov}, 

  For all infinite series occurring in (\ref{eq1})-(\ref{eq3}) a one-fold 
integral representation for arbitrary $z$ can be written \cite{semianalytic}. 
Thus these series can be continued analytically in 
the whole complex $z$-plane. 
Some of these integrals can be rewritten in terms of polylogarithms.
For example we note the following representation
\begin{eqnarray}
\sum_{n=1}^\infty \frac{1}{\left( 2n\atop n\right)} \frac{z^n}{n^a}  & = &
\int\limits_0^1 \frac{ds}{s} S_{a-2,1}\Bigl( zs(1-s) \Bigr)
\nonumber \\ & = &
\frac{1}{(a-2)!} \int\limits_0^{2\arcsin \frac{\sqrt{z}}{2}}
\Biggl[\ln z - 2 \ln \left(2 \sin \frac{\theta}{2} \right)
\Biggr]^{a-2}
   \theta\, d\theta 
\nonumber \\ & = &
- \sum_{j=0}^{a-2} \frac{(-2)^j}{(a-2-j)!j!} 
 \,( \ln z )^{a-2-j}\, \LS{j+2}{1}{2\arcsin\frac{\sqrt{z}}{2}},
\label{binomial1}
\end{eqnarray}
where  $a>1$ and $S_{a,b}(z)$ are generalized Nielsen 
polylogarithms \cite{Nielsen}.
The last two lines have been obtained in \cite{sum2}.
Substituting $z=1$ into (\ref{binomial1}) (i.e. on-shell condition for
Feynman diagram) we get 
$$
\sum_{n=1}^\infty \frac{1}{\left( 2n \atop n\right)} 
\frac{1}{n^a} \equiv V_a(\infty)
= - \frac{(-2)^{a-2}}{(a-2)!} \LS{a}{1}{\frac{\pi}{3}}\,.
$$
For $a=1,\dots,5$ we can write explicitly \cite{Lewin}
\begin{eqnarray}
\sum_{n=1}^\infty \frac{1}{\left( 2n \atop n\right)} 
 \frac{1}{n^{\{1,2,3,4,5\}}}  \z = \z
    \Biggl\{ 
     \frac13 \frac{\pi}{\sqrt{3}},\,\,\,
     \frac13 \zeta_2,\,\,\,
     \frac23 \pi \Ls{2}{\frac{\pi}{3}} - \frac{4}{3} \zeta_3,\,\,\,
     \frac{17}{36} \zeta_4,\,\,\, \nonumber\\
   \z\z  \frac{4}{9} \pi \Ls{4}{\frac{\pi}{3}}
         -\frac{19}{3} \zeta_5 - \frac{2}{3} \zeta_2 \zeta_3
     \Biggr\}.
\nonumber 
\end{eqnarray}

  However, the analytical results for arbitrary $z$ for other 
types of sums are not yet 
available. For example we may write the integral representation
\begin{eqnarray}
\z\z
\sum_{n=1}^\infty \frac{1}{\left( 2n \atop n\right)} \frac{z^n}{n^a} 
S_1(n-1) = \int\limits_0^1 \frac{ds}{s} S_{a-2,2} \Bigl( zs(1-s) \Bigr) =
\nonumber \\ 
\z\z
-\frac{2}{(a-2)!} \!\!\int\limits_0^{2\arcsin \frac{\sqrt{z}}{2}}
\Biggl[\ln z - 2 \ln \left(2 \sin \frac{\theta}{2} \right) \Biggr]^{a-2} 
\left[\Ls{2}{\pi+\theta} 
+ \theta \ln \left(2 \sin \frac{\pi+\theta}{2} \right) \right]
 \, d\theta,
\label{binomial2}
\end{eqnarray}
but we do not know how to evaluate these integrals explicitly for $a>2$ 
even at $z=1$. Nevertheless for each particular $a=1,\dots,5$ we are able 
to obtain an analytical answer for (\ref{binomial2}) at $z=1$ 
using the {\bf PSLQ} algorithm \cite{pslq}. 
This proceeds as follows. Each sum can be evaluated numerically 
with arbitrary accuracy. 
{\bf PSLQ} expresses the obtained numerical value in terms of 
given transcendental numbers. The only problem is to define 
the full set of "basis" elements. Such a basis for diagrams having two
massive particle cut was elaborated in \cite{master3}. The Ansatz for
the construction of basis up to arbitrary order have been suggested and
explicitly evaluated up to weight 5 that corresponds to
the second order in $\ep$-expansion of two-loop propagator type diagrams. 
The important role in construction of this basis belongs to Broadhurst's observations
\cite{Broadhurst3} that sixth root of unity plays an important role in
the calculation of the diagrams.  

  We have investigated all sums of 
type (\ref{binsum}) up to weight 5.
Not all of them are expressible in terms of our basis 
elements \cite{Broadhurst3,master3}
or the transcendental numbers given in \cite{massless2}. 
But it turns out that linear combinations of sums occurring in Feynman diagrams 
evaluated in the present paper are connected with basis 
given by the "sixth root of unity". All our results were obtained empirically 
by carefully compiling and examining a huge data base of high 
precision (several hundreds decimals) numerical calculations. 
Some details of this calculations are given in Appendix A.
The results of multiple binomial sum's elaboration up to weight 4 
are collected  in Appendix B. The results for sum of weight 5 are 
relatively lengthy and therefore will not be published here.  

  We note that all V-type sums (occurring e.g. in (\ref{eq2})) 
are reduced to the multiple binomial sums due to the identity

$$
\sum_{n=1}^\infty \frac{1}{n^a} \sum_{j=1}^{n-1} f (j)  = 
\sum_{n=1}^\infty f(n) \Biggl[ \zeta_a - S_a(n-1) - \frac{1}{n^a}
\Biggr].
$$

  Finally we mention that all sums occurring in Eq.(\ref{eq1}) 
are expressible in  terms of Euler-Zagier sums (\ref{euler}).

\section{Conclusion}

Collecting the results of Appendix B we obtain the following values for 
on-shell integrals
\begin{eqnarray}
&&
m^2 {\bf I_{125}} \left.\right|_{p^2=m^2} = 
\left\{ -4 \zeta_3 + 2 \pi \Ls{2}{\frac{\pi}{3}} 
+ i \pi \frac{\zeta_2}{3} 
\right\} (1+2\ep)
\nonumber \\&& 
+\ep \left(
\frac{16}{3} \Biggr[ \Ls{2}{\frac{\pi}{3}} \Biggr]^2 
- 7 \pi \Ls{3}{\frac{2\pi}{3}} - \frac{488}{9} \zeta_4
- i \pi \left\{ 
\frac{4}{9} \pi \Ls{2}{\frac{\pi}{3}}
- \frac{11}{9} \zeta_3 \right\} 
\right)
+ {\cal O}(\varepsilon^2),
\label{125}
\\ && 
m^2 {\bf I_{15}} \left.\right|_{p^2=m^2} = 
\left\{
-3 \zeta_3 + 2 \pi \Ls{2}{\frac{\pi}{3}}
+ i \pi \zeta_2
\right\} (1+2\ep)
\nonumber \\&&
+\ep \left(
6 \Biggr[ \Ls{2}{\frac{\pi}{3}} \Biggr]^2
- 9 \pi \Ls{3}{\frac{2\pi}{3}}
- \frac{2567}{36} \zeta_4 
+ i \pi \frac{5}{4} \zeta_3
\right)
+ {\cal O}(\varepsilon^2),
\label{15}
\\ && 
m^2 {\bf I_{5}} \left.\right|_{p^2=m^2} = 
\left\{
-3 \zeta_3 + i \pi \zeta_2
\right\} (1+2\ep)
\nonumber \\&&
+\ep \left(
6 \zeta_2 \ln^2 2 - \ln^4 2 - 24 \Li{4}{\frac{1}{2}}
- \frac{57}{4} \zeta_4
+ i \pi \left\{
\frac{19}{4} \zeta_3 - 9 \zeta_2 \ln 2
\right\}
\right)
+ {\cal O}(\varepsilon^2). 
\label{5}
\end{eqnarray}

  It is convenient to multiply Eqs.(\ref{125})--(\ref{5}) by $(1-2\varepsilon)$.
Then we can write (\ref{125}) and (\ref{15}) in the form
\begin{eqnarray}
&&
m^2 (1-2\ep) {\bf I} \left.\right|_{p^2=m^2} =
r_1 \zeta_3 + r_2 \pi \Ls{2}{\frac{\pi}{3}} + r_3 i \pi \zeta_2
\nonumber \\&& 
+\ep \left(
r_4 \Biggr[ \Ls{2}{\frac{\pi}{3}} \Biggr]^2 
+ r_5 \pi \Ls{3}{\frac{2\pi}{3}} 
+ r_6 \zeta_4
+ i \pi \left\{ r_7\pi \Ls{2}{\frac{\pi}{3}} + r_8 \zeta_3 \right\} 
\right)
+ {\cal O}(\varepsilon^2),
\label{result}
\end{eqnarray}
with some rational numbers $r_j$. Both $\bf I_{125}$ and $\bf I_{15}$
have a threshold at $4m^2$ plus possible thresholds at $0m^2$ and $1m^2$.
The above results suggest that all such diagrams have the form (\ref{result})
where coefficients $r_j$ depend on the distribution of the masses on lines
while the basis (\ref{result}) is defined by the topology alone.

  If a diagram has no threshold at $4m^2$ then it is expressible in terms
of Euler--Zagier sums. The three particle massive cuts lead to the 
appearance of a new structures
like $\LS{4}{1}{\frac{2\pi}{3}}$ \cite{Broadhurst3,trans,master3} 
and some others.
  
  Let us return to the results of Appendix B. One can write down a representation
in terms of a hypergeometric sum

$$
\sum_{n=1}^\infty  \frac{1}{\left( 2n \atop n\right)} \frac{z^n}{n^a} 
= \frac{z}{2}\,\,
\Fh{a+1}{a} \Fs{ \{1 \}_{a+1} }{ \frac{3}{2}, \{ 2 \}_{a-1} }{ \frac{z}{4} }.
$$

It is easy to see that multiple sums with nested harmonic summations $S_a$
can be obtained from the generating function

\begin{equation}
 {}_{p+1}F_p \left(\begin{array}{c} \{1 + a_i  \}_{p+1}; \\ 
\frac{3}{2} + b; ~ \{ 2 + c_i  \}_{p-1}; \end{array} \frac{1}{4} \right)
\label{hyp2}
\end{equation}

\noindent
by expanding (\ref{hyp2}) in powers of $a_i,c_j$ and $b$.

  There are certain sums (see Appendix B) which cannot 
be expressed (polynomially) in terms of 
a basis connected with "sixth root of unity" \cite{master3} or the one given in 
\cite{massless2}. We don't have an explanation for this phenomenon. 
However all linear combinations 
arising in the Taylor expansion of (\ref{hyp2}) are expressible in terms
of our basis.

\noindent
{\bf Acknowledgments}
We are grateful to A.~Davydychev, D.~Broadhurst,
F.~Jegerlehner and A.~Kotikov for useful
discussions and carefully reading the manuscript.
M.K's research has been supported by the DFG project FL241/4-1
and in part by RFBR $\#$98-02-16923.

\appendix

\section{Multiple precision calculation of $\Ls{n}{\theta}$
           and $\LS{n}{1}{\theta}$} 

For the purposes of {\bf PSLQ} we need to evaluate functions
${\rm Ls}_n(\theta)$ and ${\rm Ls}_n^{(1)}(\theta)$ to very high accuracy
(several hundreds of decimals). It is clear that the definitions
(\ref{ls})
are not suitable for such numerical 
calculations. For example, one needs several hours of running MAPLE 
to calculate $\Ls{6}{\pi/3}$ with accuracy about 200 decimals. 
As an alternative we found the following series
for these functions
which allows us obtain the results with needed accuracy
in a few seconds. We have ($z=4\sin^2(\theta/2)$)

\begin{eqnarray}
 {\rm Ls}_n(\theta) \z=\z (n-1)!\, \sqrt{z}\,
      \sum_{r=1}^{n} \frac{(-\log\sqrt{z})^{n-r}}{(n-r)!} 
      \,\,\sum_{k=0}^{\infty}\frac{{\left({2k}\atop{k}\right)}}{(2k+1)^r}
           \left(\frac{z}{16}\right)^k,  
\nonumber \\
{\rm Ls}_n^{(1)}(\theta) \z=\z
  - (-1)^n \frac{(n-2)!}{2^{n-2}}
   \sum_{r=0}^{n-2}\frac{(-\log z)^r}{r!} \, 
   \sum_{k=1}^{\infty}\frac{z^k}{{\left({2k}\atop{k}\right)}k^{n-r}}.
\nonumber 
\end{eqnarray}

\section{Multiple binomial sums}

  In this section we present the results of our searching of
relationships between multiple binomial sum%
\footnote{
All multiple binomial sums (\ref{binsum}) can be rewritten in
terms of function $\Psi(n)={\rm d}/{\rm d}n \log\Gamma(n)$ and its derivatives 
by means of the following relation
$$ 
\Psi^{(k-1)}(j) = (-)^k (k-1)! \left[\zeta_k - S_k(j-1) \right],
  \qquad  k>1,
$$
where $\Psi^{(k)}(z)$ is the $k$-th derivative of the $\Psi$-function.
In particular, for $k=1$ we have $\Psi(j) = S_1(j-1)-\gamma$.
}
(\ref{binsum}) up to weight 4 and the set of transcendental numbers given in
\cite{Broadhurst3,master3}.  All sums were obtained  numerically by using
multiprecision FORTRAN with accuracy of about 300 decimals and analytical
results were obtained by {\bf PSLQ}. 
The sums of weight 3 can be extracted from results 
of \cite{Broadhurst3,semianalytic}. The sums 
of weigt 4 and depths $k<3$  are investigated in  \cite{Broadhurst3}. 
 
  Below we omit argument of harmonic
sums implying that $S_a\equiv S_a(n-1)$ and $\bar{S}_a\equiv S_a(2n-1)$.
\begin{eqnarray}
\z\z
\s{{}}S_1 =
- \frac{1}{3} \frac{\pi}{\sqrt3} \ln 3 
+ \frac{4}{3} \frac{\Ls{2}{\frac{\pi}{3}}}{\sqrt{3}},
\nonumber \\ 
\z\z
\s{2}S_1 =
- \frac{4}{9} \pi \Ls{2}{\frac{\pi}{3}} + \frac{11}{9} \zeta_3,
\nonumber \\
\z\z
\s{3}S_1 =
- \pi \Ls{3}{\frac{2\pi}{3}} 
+ \frac{4}{3} \Biggr[ \Ls{2}{\frac{\pi}{3}} \Biggr]^2 
- \frac{269}{36} \zeta_4,
\nonumber \\ 
\z\z
\s{{}} \bar{S}_1 =
- \frac{1}{3}  \frac{\pi}{\sqrt{3}} \ln 3 
+ \frac{7}{3} \frac{\Ls{2}{\frac{\pi}{3}}}{\sqrt{3}}, 
\nonumber \\ 
\z\z
\s{2} \bar{S}_1 =
- \frac{7}{9} \pi \Ls{2}{\frac{\pi}{3}} + \frac{23}{9} \zeta_3,
\nonumber \\ 
\z\z
\s{3} \bar{S}_1 =
- \pi \Ls{3}{\frac{2\pi}{3}} 
+ \frac{7}{3} \Biggr[ \Ls{2}{\frac{\pi}{3}} \Biggr]^2 
- \frac{143}{18} \zeta_4, 
\nonumber \\ 
\z\z
\s{{}} S_1^2 = 
\frac{1}{3}  \frac{\pi}{\sqrt{3}} \ln^2 3 
- \frac{8}{3} \frac{\Ls{2}{\frac{\pi}{3}}}{\sqrt{3}} \ln 3 
+ \frac{55}{27}  \frac{\pi}{\sqrt{3}} \zeta_2
+ 4 \frac{\Ls{3}{\frac{2\pi}{3}}}{\sqrt{3}},
\nonumber \\ 
\z\z
\s{2} S_1^2 = 
\frac{4}{3} \pi \Ls{3}{\frac{2\pi}{3}} 
- \frac{16}{9} \Biggr[ \Ls{2}{\frac{\pi}{3}} \Biggr]^2
+ \frac{1085}{108} \zeta_4,
\nonumber \\
\z\z
\s{{}} S_1^3 =
-\frac{1}{3} \frac{\pi}{\sqrt{3}} \ln^3 3 
+ 4 \frac{\Ls{2}{\frac{\pi}{3}}}{\sqrt{3}} \ln^2 3 
- \frac{55}{9} \frac{\pi}{\sqrt{3}} \zeta_2 \ln 3 
\nonumber \\ && 
- 12 \frac{ \Ls{3}{\frac{2\pi}{3}}}{\sqrt{3}}\ln 3
+ \frac{2}{3} \zeta_2 \frac{\Ls{2}{\frac{\pi}{3}}}{\sqrt{3}}  
- \frac{179}{27}  \frac{\pi}{\sqrt{3}} \zeta_3
- \frac{92}{27} \frac{\Ls{4}{\frac{\pi}{3}}}{\sqrt{3}}
+ 8 \frac{\Ls{4}{\frac{2\pi}{3}}}{\sqrt{3}},
\nonumber \\ 
\z\z
\s{{}} S_1\bar{S}_1 =
\frac{1}{3} \frac{\pi}{\sqrt{3}} \ln^2 3
- \frac{11}{3} \frac{\Ls{2}{\frac{\pi}{3}}}{\sqrt{3}} \ln 3 
+ \frac{157}{54} \frac{\pi}{\sqrt{3}} \zeta_2 
+ \frac{11}{2}  \frac{\Ls{3}{\frac{2\pi}{3}}}{\sqrt{3}},
\nonumber \\ 
\z\z
\s{2} S_1\bar{S}_1 =
\frac{11}{6} \pi \Ls{3}{\frac{2\pi}{3}} 
- \frac{28}{9} \Biggr[ \Ls{2}{\frac{\pi}{3}} \Biggr]^2 
+ \frac{3125}{216}\zeta_4, 
\nonumber \\ 
\z\z
\s{{}} S_1^2\bar{S}_1 =
-\frac{1}{3} \frac{\pi}{\sqrt{3}} \ln^3 3
+ 5 \frac{\Ls{2}{\frac{\pi}{3}}}{\sqrt{3}} \ln^2 3
- \frac{212}{27} \frac{\pi}{\sqrt{3}} \zeta_2 \ln 3
\nonumber \\ &&
- 15 \frac{\Ls{3}{\frac{2 \pi}{3}}}{\sqrt{3}} \ln 3
+ \frac{8}{9} \zeta_2 \frac{\Ls{2}{\frac{\pi}{3}}}{\sqrt{3}}
- \frac{727}{81} \frac{\pi}{\sqrt{3}} \zeta_3
- \frac{298}{81} \frac{\Ls{4}{\frac{\pi}{3}}}{\sqrt{3}}
+  10 \frac{\Ls{4}{\frac{2 \pi}{3}}}{\sqrt{3}},
\nonumber \\ 
\z\z
\s{{}} S_2 = 
\frac{1}{27} \frac{\pi}{\sqrt{3}} \zeta_2,
\nonumber \\
\z\z
\s{2} S_2 =  
\frac{5}{108} \zeta_4,
\nonumber \\ 
\z\z
\s{{}} S_3 =  
- \frac{2}{3} \zeta_2 \frac{\Ls{2}{\frac{\pi}{3}}}{\sqrt{3}} 
- \frac{16}{9} \frac{\pi}{\sqrt{3}} \zeta_3
+ \frac{4}{3} \frac{\Ls{4}{\frac{\pi}{3}}}{\sqrt{3}},
\nonumber \\ && 
\nonumber \\ 
\z\z
\s{{}} S_1 S_2 =  
- \frac{1}{27} \frac{\pi}{\sqrt{3}} \zeta_2 \ln3 
- \frac{2}{9} \zeta_2 \frac{\Ls{2}{\frac{\pi}{3}}}{\sqrt{3}}
+ \frac{49}{81}  \frac{\pi}{\sqrt{3}} \zeta_3 
- \frac{20}{81} \frac{\Ls{4}{\frac{\pi}{3}}}{\sqrt{3}},
\nonumber \\
\z\z
\s{{}} S_2\bar{S}_1 =  
-\frac{1}{27} \frac{\pi}{\sqrt{3}}\zeta_2 \ln 3
+ \frac{4}{9} \zeta_2 \frac{\Ls{2}{\frac{\pi}{3}}}{\sqrt{3}}
+ \frac{229}{81} \frac{\pi}{\sqrt{3}} \zeta_3
- \frac{146}{81} \frac{\Ls{4}{\frac{\pi}{3}}}{\sqrt{3}},
\nonumber \\ 
\z\z
\s{{}} \left( \bar{S}_1^2 + \bar{S}_2 \right) = 
\frac{1}{3} \frac{\pi}{\sqrt{3}} \ln^2 3
- \frac{14}{3} \frac{\Ls{2}{\frac{\pi}{3}}}{\sqrt{3}} \ln 3
+ \frac{113}{27} \frac{\pi}{\sqrt{3}} \zeta_2
+ 7 \frac{\Ls{3}{\frac{2\pi}{3}}}{\sqrt{3}},
\nonumber \\ 
\z\z
\s{2} \left( \bar{S}_1^2 + \bar{S}_2 \right) = 
\frac{7}{3} \pi \Ls{3}{\frac{2\pi}{3}}
- \frac{49}{9} \Biggl[ \Ls{2}{\frac{\pi}{3}} \Biggr]^2
+ \frac{4505}{216} \zeta_4,
\nonumber \\ 
\z\z
\s{{}} S_1 \left( \bar{S}_1^2 + \bar{S}_2 \right) = 
-\frac{1}{3}\frac{\pi}{\sqrt{3}} \ln^3 3
+ 6 \frac{\Ls{2}{\frac{\pi}{3}}}{\sqrt{3}} \ln^2 3
- 10\frac{\pi}{\sqrt{3}} \zeta_2 \ln 3
- \frac{112}{9} \frac{\pi}{\sqrt{3}} \zeta_3
\nonumber \\ &&
- 18 \frac{\Ls{3}{\frac{2\pi}{3}}}{\sqrt{3}} \ln 3
+ \frac{2}{3} \zeta_2 \frac{\Ls{2}{\frac{\pi}{3}}}{\sqrt{3}} 
- \frac{8}{3} \frac{\Ls{4}{\frac{\pi}{3}}}{\sqrt{3}}
+ 12 \frac{\Ls{4}{\frac{2 \pi}{3}}}{\sqrt{3}},
\nonumber \\ 
\z\z
\s{{}} \left( \bar{S}_1^3 + 3\bar{S}_1\bar{S}_2 + 2\bar{S}_3 \right) = 
-\frac{1}{3}\frac{\pi}{\sqrt{3}} \ln^3 3
+ 7 \frac{\Ls{2}{\frac{\pi}{3}}}{\sqrt{3}} \ln^2 3
- \frac{394}{27} \frac{\pi}{\sqrt{3}} \zeta_3
\nonumber \\ &&
- \frac{113}{9} \frac{\pi}{\sqrt{3}} \zeta_2 \ln 3
- 21 \frac{\Ls{3}{\frac{2\pi}{3}}}{\sqrt{3}} \ln 3
+ \frac{2}{3} \zeta_2 \frac{\Ls{2}{\frac{\pi}{3}}}{\sqrt{3}}
- \frac{28}{27}  \frac{\Ls{4}{\frac{\pi}{3}}}{\sqrt{3}}
+ 14 \frac{\Ls{4}{\frac{2 \pi}{3}}}{\sqrt{3}}.
\nonumber 
\end{eqnarray}


\end{document}